\documentclass[aps, showpacs, floatfix, twocolumn, reprint, 
twoside, prc, preprintnumbers, 
superscriptaddress]{revtex4-1}
\usepackage{array}  % for emphasizing table rows/columns
\usepackage{graphicx}  % include figure files
\usepackage[export]{adjustbox}  % for figure alignment 
\usepackage{float}
\usepackage{epstopdf}
\usepackage{amsmath}
\usepackage{amssymb}
\usepackage{bm}
\usepackage{xcolor,xspace}
\usepackage{multirow}
\usepackage{mhchem}  % chemistry package for isotope 
\usepackage{hyperref}
\hypersetup{colorlinks=True,urlcolor=blue,linkcolor=blue,
citecolor=blue,filecolor=black}
\usepackage{braket}
\usepackage{wasysym}  % for the use of \CIRCLE
\usepackage{scrextend}  % to use \footref - refer to the 
\usepackage[percent]{overpic}  % insert LaTex text over 
\usepackage[capitalise]{cleveref}  % to insert multiple 
\colorlet{Changes@Color}{red}  % changes in red color
\newcommand{\ba}{\begin{eqnarray}}
\newcommand{\ea}{\end{eqnarray}}
\newcommand{\bsub}{\begin{subequations}}
\newcommand{\esub}{\end{subequations}}

\begin{document}
\title{Mixed configurations and intertwined quantum phase 
transitions in odd-mass nuclei}
\author{N.~Gavrielov}\email{noam.gavrielov@yale.edu}
\affiliation{Center for Theoretical Physics, Sloane Physics 
Laboratory, Yale University, New Haven, Connecticut 
06520-8120, USA}
\affiliation{Racah Institute of Physics, The Hebrew 
University, Jerusalem 91904, Israel}
\author{A.~Leviatan}\email{ami@phys.huji.ac.il}
\affiliation{Racah Institute of Physics, The Hebrew 
University, 
Jerusalem 91904, Israel}
\author{F.~Iachello}\email{francesco.iachello@yale.edu}
\affiliation{Center for Theoretical Physics, Sloane Physics 
Laboratory, 
Yale University, New Haven, Connecticut 06520-8120, USA}

\date{\today}  
\begin{abstract}
  We introduce a new Bose-Fermi framework for studying 
  spectral properties and quantum phase transitions (QPTs) 
  in odd-mass nuclei, in the presence of configuration 
  mixing. A detailed analysis of odd-mass Nb isotopes 
  discloses the effects of an abrupt crossing of states in
  normal and intruder configurations (Type~II QPT), 
  accompanied by a gradual evolution from spherical- to 
  deformed-core shapes within the intruder configuration 
  (Type~I QPT). The pronounced presence of both types of 
  QPTs demonstrates, for the first time, the occurrence of
  intertwined QPTs in odd-mass nuclei.
\end{abstract}

\maketitle

Structural changes induced by variation of parameters in 
the Hamiltonian, called quantum phase transitions 
(QPTs)~\cite{Gilmore1978a,Gilmore1979}, are salient 
phenomena in dynamical systems, and form the subject of 
ongoing intense experimental and theoretical activity in 
diverse fields~\cite{carr2010QPT}. In nuclear physics, most 
of the attention has been devoted to the evolution of 
structure with nucleon number, exhibiting two types of 
phase transitions. The first, denoted as 
\mbox{Type~I}~\cite{Dieperink1980}, is a shape-phase 
transition in a single configuration, as encountered in the 
neutron number~90 region~\cite{Cejnar2010}. The second, 
denoted as \mbox{Type~II}~\cite{Frank2006}, is a phase 
transition involving an abrupt crossing of different 
configurations, as encountered in nuclei near (sub-) shell 
closure~\cite{Heyde11}. If the mixing is small, the Type~II 
QPT can be accompanied by a distinguished Type~I QPT within 
each configuration separately. Such a scenario, referred to 
as intertwined QPTs (IQPTs), was recently shown to occur in 
the Zr isotopes~\cite{Gavrielov2019,Gavrielov2022}.

Most studies of QPTs in nuclei have focused on systems with 
even numbers of protons and neutrons ~\cite{Cejnar2010, 
Heyde11, Casten2009, Iachello2011, Fortunato2021}. The 
structure of odd-mass nuclei is more complex due to the  
presence of both collective and single-particle degrees of 
freedom. Consequently, QPTs in such nuclei have been far 
less studied. Fully microscopic approaches to QPTs in 
medium-heavy odd-mass nuclei, such as the large-scale shell 
model~\cite{SM2005} and beyond-mean-field 
methods~\cite{BMF2014}, are computationally demanding and 
encounter difficulties. Alternative approaches have been 
proposed, including algebraic modeling (shell-model 
inspired~\cite{ScholtenBlasi1982, IBFMBook, Petrellis2011a}
and symmetry-based~\cite{Jolie2004, Alonso2005, Alonso2007, 
Alonso2009, Boyukata2010, Petrellis2011b, Boyukata2021}) 
and density functionals-based mean-field 
methods~\cite{Nomura2016a, Nomura2016b, Nomura2020, 
QuanMeng2018}, involving particle-core coupling schemes with
boson-fermion or collective Hamiltonians. So far these 
approaches were restricted to Type~I QPTs in odd-mass 
nuclei without configuration mixing.

The goals of the present Letter are twofold.
(\textit{i})~To~introduce a framework for studying spectral 
properties and QPTs with configuration mixing, in odd-mass 
nuclei. This is motivated by a wealth of new experimental 
data on shape-coexisting states in such nuclei near shell 
closure~\cite{Spagnoletti2019, Boulay2020}, whose awaiting 
interpretation necessitates multiple configurations.
(\textit{ii})~To apply the formalism and show evidence for
concurrent types of QPTs exemplifying, for the first time,
IQPTs in odd-mass nuclei.

Odd-$A$ nuclei are treated in the interacting boson-fermion
model (IBFM)~\cite{IBFMBook}, as a system of monopole ($s$) 
and quadrupole ($d$) bosons, representing valence nucleon 
pairs, and a single (unpaired) nucleon. We propose to 
extend the IBFM to include core excitations and obtain a 
boson-fermion model with configuration mixing (IBFM-CM), 
employing a Hamiltonian of the form,
\ba
  \label{eq:ham}
\hat H = \hat H_{\rm b} + \hat H_{\rm f} + \hat V_{\rm bf} 
~.
\ea
The boson part ($\hat H_{\rm b}$) is the Hamiltonian of the 
configuration mixing model (IBM-CM) of~\cite{Duval1981, 
Duval1982}. For two configurations (A,B), it can be cast in 
matrix form~\cite{Frank2006},
\begin{equation}\label{eq:H_B}
\hat H_{\rm b} = \begin{bmatrix}
  \hat H_{\rm b}^{\rm A}(\xi^{(\rm A)}) &
\hat{W}_{\rm b}(\omega)\\
\hat{W}_{\rm b}(\omega) 
& \hat H_{\rm b}^{\rm B}(\xi^{(\rm B)})
\end{bmatrix} ~.
\end{equation}
Here, $\hat H_{\rm b}^{\rm A}(\xi^{(\rm A)})$ represents 
the normal A configuration ($N$ boson space) and $\hat 
H_{\rm b}^{\rm B}(\xi^{(\rm B)})$ represents the intruder 
B~configuration ($N\!+\!2$ boson space), corresponding to 
2p-2h excitations across the (sub-) shell closure. Standard 
forms, as in Eq.~(4) of Ref.~\cite{Gavrielov2022}, include 
pairing, quadrupole, and rotational terms, and a mixing term
$\hat W_{\rm b} \!=\! \omega [(d^\dag d^\dag)^{(0)}
  \!+\! (s^\dag)^2] +\text{Hermitian conjugate (H.c.)}$.
Such IBM-CM Hamiltonians have been used extensively for the
study of configuration-mixed QPTs and shape coexistence
in even-even nuclei~\cite{Duval1981, Duval1982, 
Sambataro1982, Ramos2014, Ramos2015, Nomura2016c, Lev2018, 
Ramos2022, Ramos2019, Gavrielov2019, Gavrielov2022}.

The fermion Hamiltonian ($\hat H_{\rm f}$) of 
Eq.~(\ref{eq:ham}) has the form
\begin{equation}\label{eq:H_F}
\hat H_{\rm f} = \begin{bmatrix}
\sum_j\epsilon^{(\rm A)}_j \hat n_j & 0 \\
0 & \sum_j\epsilon^{(\rm B)}_j \hat n_j
\end{bmatrix} ~,
\end{equation}
where $j$ is the angular momentum of the occupied orbit, 
$\hat n_j$ the corresponding number operator, and 
$\epsilon^{(\rm i)}_j\,({i=\rm A,B)}$ are the 
single-particle energies for each configuration. The 
boson-fermion 
interaction has the form
\begin{equation}\label{eq:V_BF}
\hat V_{\rm bf} = \begin{bmatrix}
  \hat V^{\rm A}_{\rm bf}(\zeta^{(\rm A)}) &
 \hat{W}_{\rm bf}(\omega_j)\\
\hat{W}_{\rm bf}(\omega_j) &
\hat V^{\rm B}_{\rm bf}(\zeta^{(\rm B)})
\end{bmatrix} ~.
\end{equation}
Here, $\hat V^{(i)}_{\rm bf}\,({i=\rm A,B})$ involve 
monopole, quadrupole, and exchange terms with parameters 
$\zeta^{(i)} \!=\! 
(A^{(i)}_{j},\Gamma^{(i)}_{jj'},\Lambda^{(i)j''}_{jj'})$.
Using the microscopic interpretation of the 
IBFM~\cite{IBFMBook}, these couplings can be expressed in 
terms of strengths 
($A^{(i)}_{0},\Gamma^{(i)}_{0},\Lambda^{(i)}_{0}$) and
occupation probabilities $(u_j,v_j)$. The term 
$\hat{W}_{\rm bf}(\omega_j) \!=\! 
\sum_j\omega_j \hat n_j [(d^{\dag}d^{\dag})^{(0)}
\!+\! (s^{\dag})^2 
+ \text{H.c.}]$, controls the mixing for each orbit.

The Hamiltonian of Eq.~(\ref{eq:ham}) is diagonalized 
numerically. The resulting eigenstates $\ket{\Psi;J}$ are 
linear combinations of wave functions $\Psi_{\rm A}$ 
and $\Psi_{\rm B}$, involving bosonic basis states in the 
two spaces $\ket{[N],\alpha,L}$ and $\ket{[N+2],\alpha,L}$.
The boson ($L$) and fermion ($j$) angular momenta are 
coupled to $J$, $\ket{\Psi;J} \!=\! 
\sum_{\alpha,L,j}C^{(N,J)}_{\alpha,L,j}
  \ket{\Psi_{\rm A};[N],\alpha,L;j;J}
  + \sum_{\alpha,L_,j}C^{(N+2,J)}_{L,j}
  \ket{\Psi_{\rm B};[N+2],\alpha,L;j;J}$.
The probability of normal-intruder mixing is given by
\begin{equation}\label{eq:norm_int}
  a^2\!=\!\sum_{\alpha,L,j}|C^{(N,J)}_{\alpha,L,j}|^2,
  \;\;
  b^2\!=\!1-a^2
  \!=\!\sum_{\alpha,L,j}|C^{(N+2,J)}_{\alpha,L,j}|^2.
\end{equation}
Operators inducing electromagnetic transitions of type
$\sigma$ and multipolarity $L$ contain boson and fermion
parts,
\ba
\hat{T}(\sigma L) =
\hat{T}_{\rm b}(\sigma L) + \hat T_{\rm f}(\sigma L) ~.
\label{TsigL}
\ea
For $E2$ transitions, $\hat{T}_{\rm b}(E2) \!=\!
e^{(\rm{A})}\hat Q^{(N)}_{\chi} + e^{(\rm{B})}\hat 
Q^{(N+2)}_{\chi}$,
where the superscript $(N)$ denotes a projection onto the 
$[N]$ boson space and $\hat Q_\chi \!=\! d^\dag s+s^\dag 
\tilde d\!+\! \chi (d^\dag \tilde d)^{(2)}$.
For $M1$ transitions, $\hat T_{\rm b}(M1)\!=\! 
\sum_{i}\sqrt{\frac{3}{4\pi}}g^{(i)} \hat L^{(N_i)}\!+\!
\tilde{g}^{(i)}[\hat Q^{(N_i)}_{\chi} \times \hat 
L^{(N_i)}]^{(1)}$, where $i\!=\!({\rm A,B})$, $N_{\rm 
A}\!=\!N,\, N_{\rm B}\!=\!N\!+\!2$. The fermion operators 
$\hat T_{\rm f}(\sigma L)$ have the standard 
form~\cite{IBFMBook} with effective charge $e_f$ for $E2$ 
transitions, and $g_s$ quenched by 20\% for $M1$ 
transitions. In what follows, we apply the above IBFM-CM 
framework to the study of QPTs in the odd-mass Nb isotopes.

The $\ce{_{41}^ANb}$ isotopes with mass number 
\mbox{$A\!=\!\text{93--105}$} are described by coupling a 
proton to their respective $\ce{_{40}Zr}$ cores with 
neutron number 52--64. In the latter, the normal 
A~configuration corresponds to having no active protons 
above the $Z\!=\!40$ subshell gap, and the intruder 
B~configuration corresponds to two-proton excitation from 
below to above this gap, creating 2p-2h states. The 
parameters of $\hat H_{\rm b}$~\eqref{eq:H_B} and boson 
numbers are taken to be the same as in a previous 
calculation of these Zr isotopes (see Table~V of 
Ref.~\cite{Gavrielov2022}), except for $\chi\!=\!-0.565$ 
at neutron number 64.

For $\ce{_{41}Nb}$ isotopes, the valence protons reside in 
the $Z\!=\!\text{28--50}$ shell. Using as an input the 
empirical single-proton energies (taken from Table XI of 
Ref.~\cite{Barea2009}) and a pairing gap $\Delta_{\rm 
F}\!=\!1.5$~MeV, a BCS calculation yields the single 
quasiparticle energies ($\epsilon_j$) and occupation 
probabilities ($v^2_j$) for the considered $1g_{9/2}, 
2p_{1/2}, 2p_{3/2}, 1f_{5/2}$ orbits, assuming, for 
simplicity, the same parameters for both configurations. 
The derived $\epsilon_j$ and $v^2_j$, and the common 
strengths $(A_0,\Gamma_0,\Lambda_0)$, obtained by a fit, 
are listed in \cref{tab:parameters}. As seen, the monopole 
term ($A_0$) vanishes for neutron number 52--56 and 
corrects the quasiparticle energies at neutron number  
58--64. The quadrupole term ($\Gamma_0$) is constant for  
the entire chain. The exchange term ($\Lambda_0$) increases 
towards the neutron midshell~\cite{IBFMBook}. Altogether, 
the values of the parameters are either constant for the 
entire chain or segments of it and vary smoothly. We take  
$\omega_j\!=\!0$ in the $\hat W_{\rm bf}$ term of 
Eq.~(\ref{eq:V_BF}), since for equal $\omega_j$ it 
coincides with the $\hat W_{\rm b}$ term of 
Eq.~(\ref{eq:H_B}).
\begin{table}[t]
\begin{center}
\caption{\label{tab:parameters}
\small
Parameters in MeV of the boson-fermion interactions,
$\hat V^{(i)}_{\rm bf}$ of \cref{eq:V_BF}, obtained from a 
fit assuming $A^{(i)}_0\!=\!A_0$, 
$\Gamma^{(i)}_0\!=\!\Gamma_0$ and 
$\Lambda^{(i)}_0\!=\!\Lambda_0$, 
$\epsilon^{(i)}_j\!=\!\epsilon_j$, where $({i=\rm A,B})$. 
From a BCS calculation, $\epsilon_j\!=\!1.639$, 1.524, 
2.148 and 2.519~MeV  and $v^2_j\!=\!0.299$, 0.589, 0.858, 
and 0.902, for the $1g_{9/2},2p_{1/2},2p_{3/2},1f_{5/2}$ 
orbits, respectively, and Fermi energy  
$\lambda_F\!=\!2.024$~MeV.}
\begin{tabular}{cccccccc}
\hline
Neutron number & $52$ & $54$  & $56$ &  $58$ & $60$ & $62$ 
& $64$ \\
\hline
$A_0$      &0.00&0.00&0.00&$-0.11$&$-0.20$&$-0.20$&$-0.20$\\
$\Gamma_0$ &1.00&1.00&1.00&1.00   &  1.00 &  1.00 &  1.00 \\
$\Lambda_0$&1.00&1.00&3.00&3.00   &  3.80 &  3.80 &  3.80 \\
\hline 
\end{tabular}
\end{center}
\end{table}

\begin{figure}[t!]
\centering
\includegraphics[width=1\linewidth]{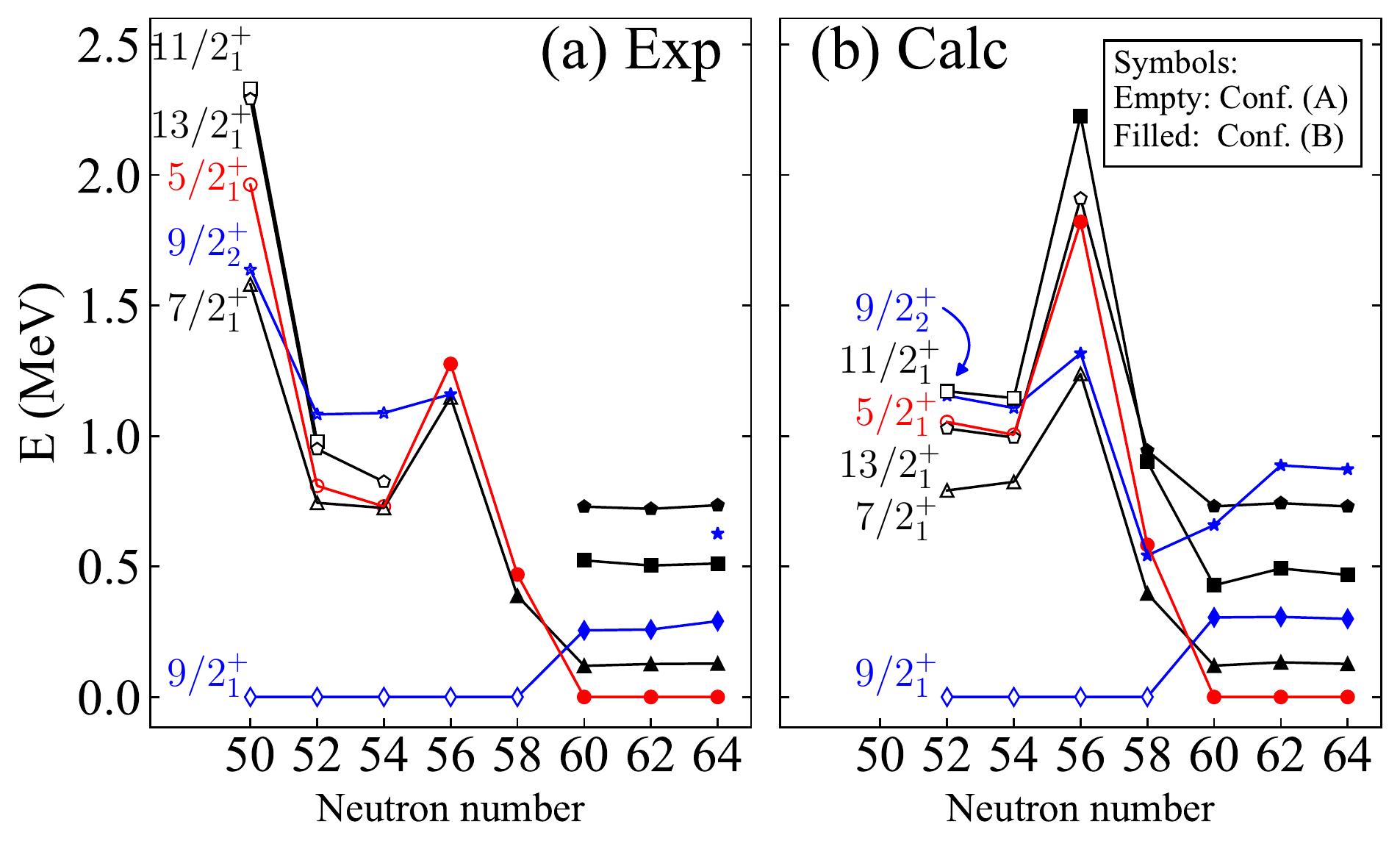} %
\caption{Comparison between 
(a)~experimental~\cite{NDS.114.1293.2013, 
NDS.112.1163.2011, NDS.111.2555.2010, NDS.111.525.2010, 
NDS.145.25.2017} and (b)~calculated lowest-energy   
positive-parity levels in Nb isotopes. Open (solid) 
symbols indicate a state dominated by the normal 
A~configuration (intruder B~configuration), with   
assignments based on \cref{eq:norm_int}. In particular,   
the $9/2^+_1$ state is in the A (B) configuration for  
neutron number 52--58 (60--64) and the $5/2^+_1$ state is  
in the A (B) configuration for 52--54 (56--64). Note that 
the calculated values start at 52, while the experimental 
values include the closed shell at 50. 
\label{fig:energies-p}}
\end{figure}

In the present Letter, we concentrate on the 
positive-parity states in Nb isotopes, postponing a 
discussion of both parity states to a longer paper. Such a 
case reduces to a single-$j$ calculation, with the 
$\pi(1g_{9/2})$ orbit coupled to the boson core. 
Figure \ref{fig:energies-p} shows the experimental and 
calculated levels of selected states, along with 
assignments to configurations based on 
Eq.~(\ref{eq:norm_int}). Open (solid) symbols indicate a 
dominantly normal (intruder) state with small (large) $b^2$ 
probability. In the region between neutron number 50 and 
56, there appear to be two sets of levels with a weakly 
deformed structure, associated with configurations A and B. 
All levels decrease in energy for 52--54, away from the 
closed shell, and rise again at 56 due to the 
$\nu(2d_{5/2})$ subshell closure. From 58, there is a 
pronounced drop in energy for the states of the 
B~configuration. At 60, the two configurations cross, 
indicating a Type~II QPT, and the ground state changes from 
$9/2^+_1$ to $5/2^+_1$, becoming the bandhead of a 
$K=5/2^+$ rotational band composed of $5/2^+_1, 7/2^+_1, 
9/2^+_1, 11/2^+_1, 13/2^+_1$ states. The intruder 
B~configuration remains strongly deformed and the band 
structure persists beyond 60. The above trend is similar to 
that encountered in the even-even Zr isotopes with the same 
neutron numbers (see Fig.~14 of Ref.~\cite{Gavrielov2022}).

A possible change in the angular momentum of the ground 
state ($J^{+}_{\rm{gs}}$) is a characteristic signature of 
Type~II QPTs in odd-mass, unlike even-even nuclei where the 
ground state remains $0^+$ after the crossing. It is an 
important measure for the quality of the calculations, 
since a mean-field approach, without configuration mixing, 
fails to reproduce the switch $9/2^+_1\!\to\!5/2^+_1$ in 
$J^{+}_{\rm{gs}}$ for the Nb isotopes~\cite{Guzman2011}. 
Figure 2(a) shows the percentage of the wave function 
within the B~configuration for $J^+_{\rm{gs}}$ and 
$7/2^+_1$, as a function of neutron number across the Nb 
chain. The rapid change in structure of $J^+_{\rm{gs}}$ 
from the normal A~configuration in $^{93-99}$Nb (small 
$b^2$ probability), to the intruder B~configuration in 
$^{101-105}$Nb (large $b^2$) is clearly evident, signaling 
a Type~II QPT. The configuration change appears sooner in 
the $7/2^+_1$ state, which changes to the B~configuration 
already in $^{99}$Nb. Outside a narrow region near neutron 
number 60, where the crossing occurs, the two 
configurations are weakly mixed and the states retain a 
high level of purity. Such a trend is similar to that 
encountered for the $0^+_1$ and $2^+_1$ states in the 
respective $\ce{_{40}Zr}$ cores (see Fig.~10 of 
Ref.~\cite{Gavrielov2022}).
\begin{figure}[t]
\centering
\includegraphics[width=1\linewidth]{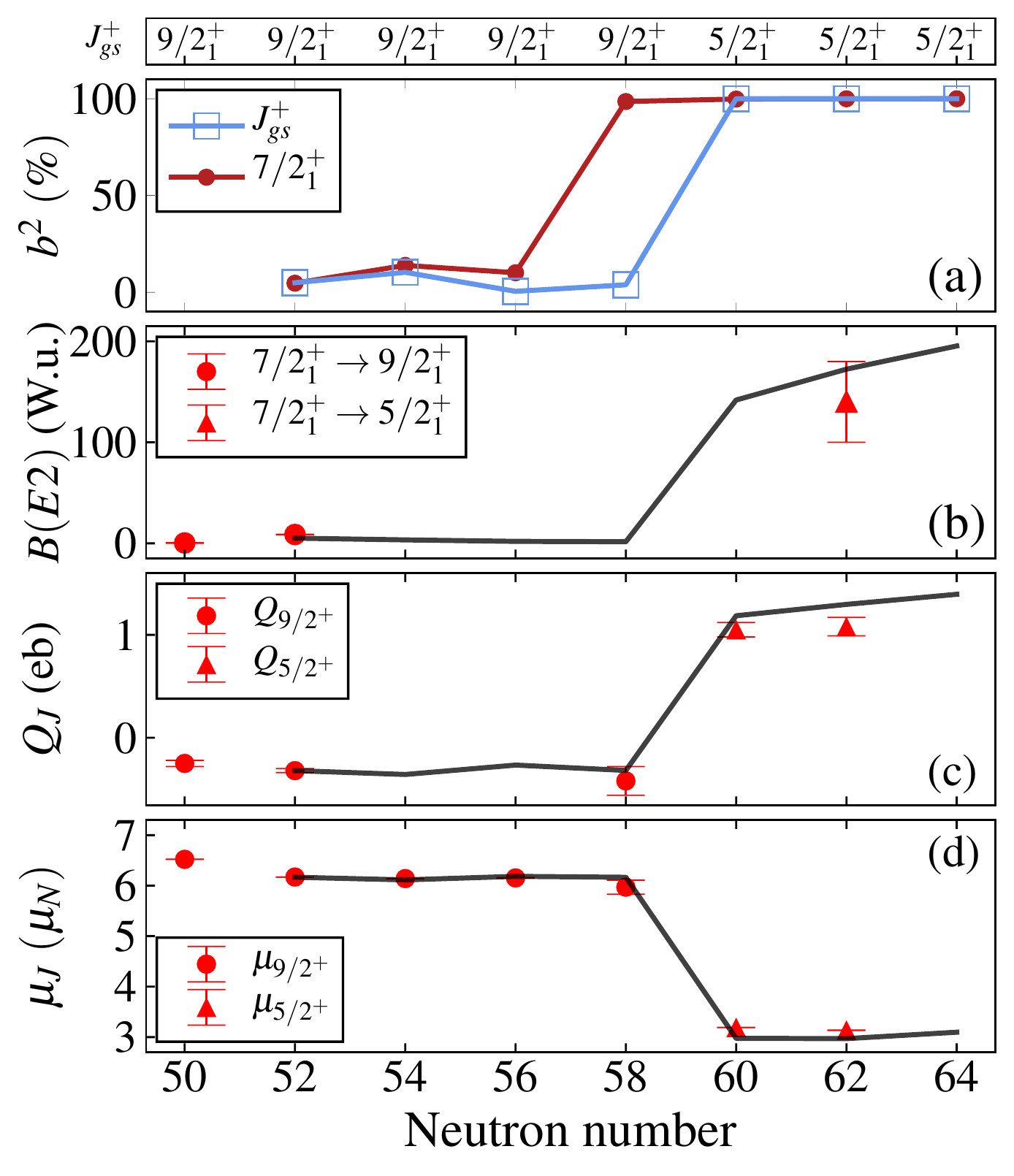} %
\caption{Evolution of spectral properties along the Nb 
chain. Symbols (solid lines) denote experimental data 
(calculated results). (a)~Percentage of the intruder (B) 
component [the $b^2$ probability in \cref{eq:norm_int}], in 
the ground state ($J^+_{\rm{gs}}$) and the first-excited 
positive-parity state ($7/2^+_1$) of $^{93-103}$Nb. The 
values of $J^+_{\rm{gs}}$ are indicated at the top. 
(b)~$B(E2; 7/2^+_1\to J^{+}_{\rm{gs}})$ in Weisskopf units 
(W.u.) (c)~Quadrupole moments of $J^+_{\rm{gs}}$ in $eb$. 
(d)~Magnetic moments of $J^+_{\rm{gs}}$ in $\mu_N$. Data in 
(b)--(d), are taken from Refs.~\cite{NDS.114.1293.2013, 
NDS.112.1163.2011, NDS.110.2081.2009}, 
Refs.~\cite{NDS.114.1293.2013, NDS.112.1163.2011, 
Cheal2009}, and Refs.~\cite{NDS.114.1293.2013, 
NDS.112.1163.2011, NDS.111.2555.2010, NDS.111.525.2010, 
NDS.145.25.2017, Cheal2009}, respectively. 
\label{fig:order-params}}
\end{figure}
\begin{figure*}
\centering
\includegraphics[width=1\linewidth]{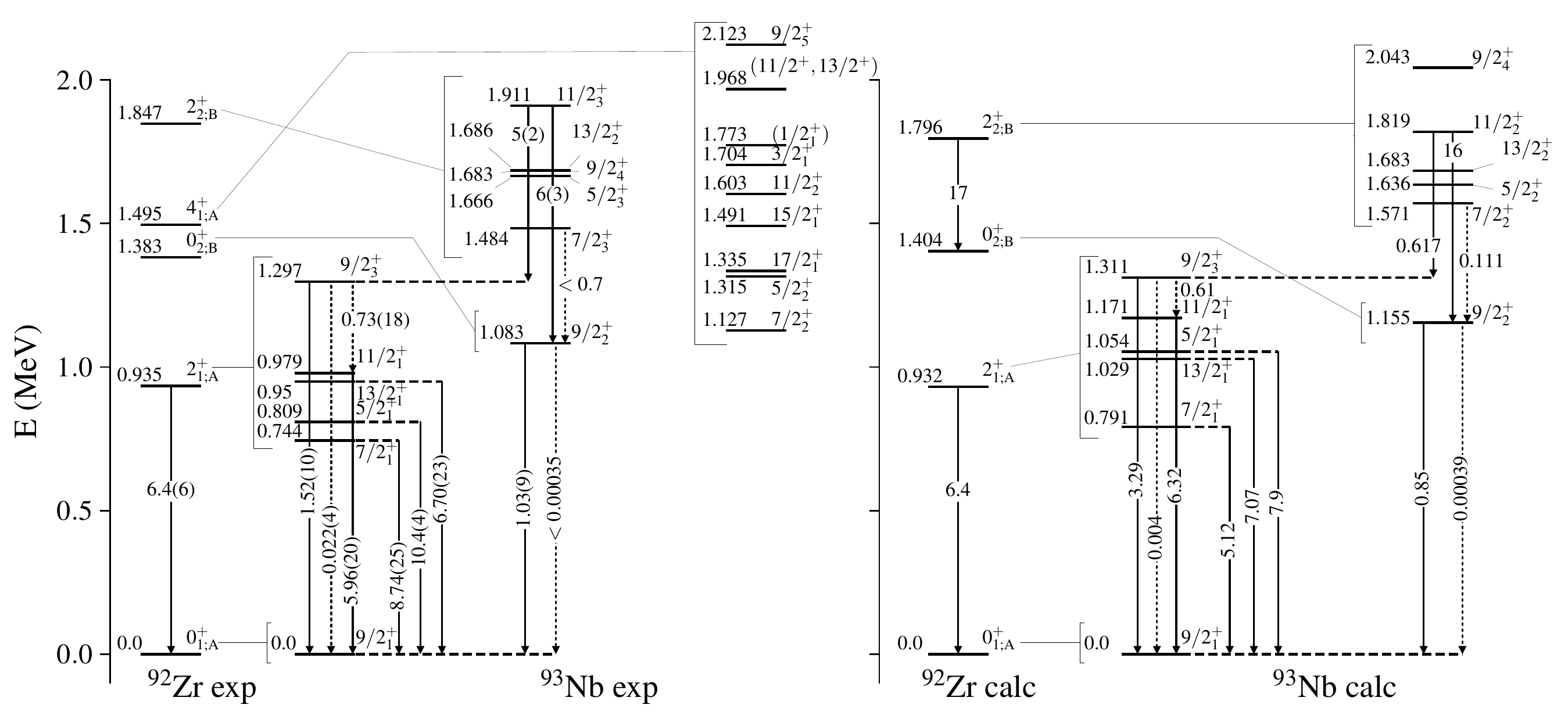}
\caption{Experimental (left) and calculated (right) energy 
levels in MeV, and $E2$ (solid arrows) and $M1$ (dashed 
arrows) transition rates in W.u., for $^{93}$Nb and 
$^{92}$Zr. Lines connect $L$ levels in $^{92}$Zr to sets of 
$J$ levels in $^{93}$Nb, indicating the weak coupling 
$(L\otimes \tfrac{9}{2})J$. Data taken from 
Refs.~\cite{Orce2010, NDS.112.1163.2011}. Note that the 
observed $4^+_{1;\rm A}$ state in $^{92}$Zr is outside the 
$N\!=\!1$ model space.\label{fig:93Nb-p}}
\end{figure*}
\begin{figure}
\centering
\includegraphics[width=1\linewidth]{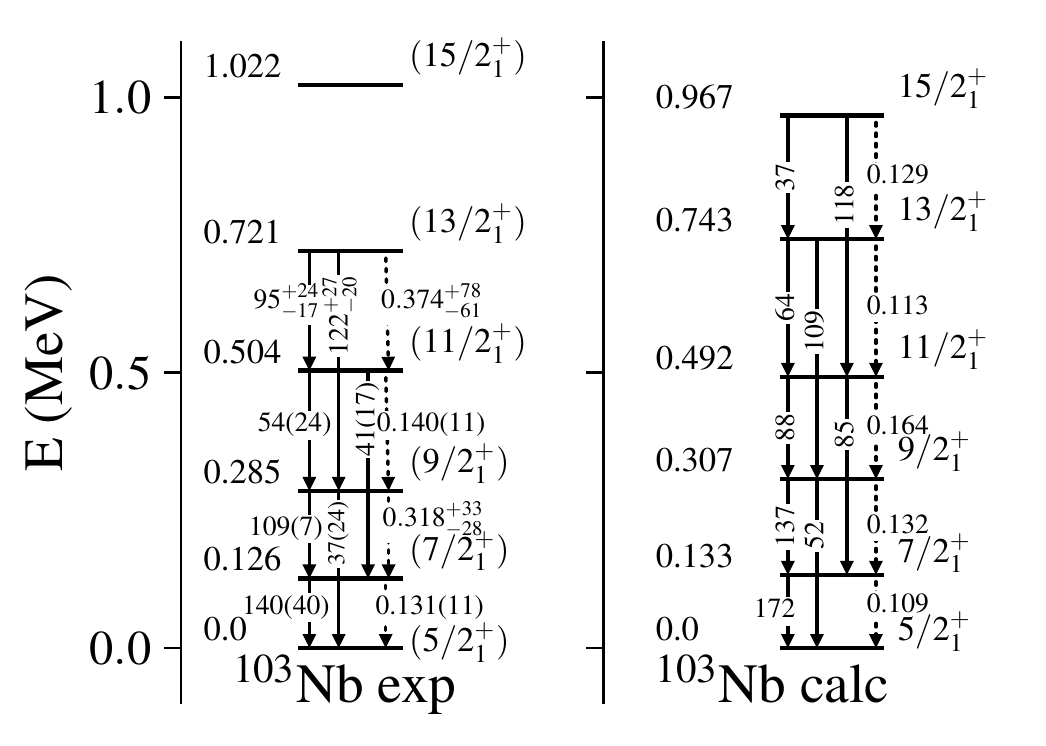} %
\caption{Experimental (left) and calculated (right) energy 
levels in MeV, and $E2$ (solid arrows) and $M1$ (dashed 
arrows) transition rates in W.u., for $\ce{^{103}Nb}$. Data 
taken from Refs.~\cite{NDS.110.2081.2009, Hagen2017}. 
\label{fig:103Nb-p}}
\end{figure}

Electromagnetic transitions and moments provide further 
insight into the nature of QPTs. For $\hat{T}_{\rm b}(E2)$ 
of Eq.~(\ref{TsigL}), we adopt the same parameters 
$(e^{(\rm A)},e^{(\rm B)},\chi)$ used for the core Zr 
isotopes~\cite{Gavrielov2022}, with a slight modification of
\mbox{$e^{(\rm A)}\!=\!2.45,\,1.3375~\sqrt{\text{W.u.}}$} 
for neutron numbers 52--54 and 
\mbox{$e^{(\rm B)}\!=\!2.0325~\sqrt{\text{W.u.}}$} for 62. 
The fermion effective charge in $\hat{T}_{\rm f}(E2)$ is 
\mbox{$e_f\!=\!-2.361~\sqrt{\text{W.u.}}$}, determined from 
a fit to the ground state quadrupole moment of $^{93}$Nb. 
For $\hat{T}_{\rm b}(M1)$ we use $g^{(\rm 
A)}\!=\!-0.21,\,-0.42\mu_N$ for neutron number 52--54 and 
zero otherwise, $g^{(\rm B)}\!=\!(Z/A)\mu_N$ and 
$\tilde{g}^{(\rm A)}\!=\!\tilde{g}^{(\rm B)}\!=\!0\; 
(-0.017\mu_N)$ for 52--56 (58--64). For $\hat{T}_{\rm 
f}(M1)$ we use $g_{\ell}\!=\!1\mu_N$ and 
$g_{s}\!=\!4.422\mu_N$.

The $B(E2; 7/2^+_1\to J^{+}_{\rm{gs}})$ and quadrupole 
moment of $J^{+}_{\rm{gs}}$ are shown in 
Fig.~\ref{fig:order-params}(b) and 
Fig.~\ref{fig:order-params}(c), respectively. These 
observables are related to the deformation, the order 
parameter of the QPT. Although the data are incomplete, one 
can still observe small (large) values of these observables 
below (above) neutron number 60, indicating an increase in 
deformation. The calculation reproduces well this trend and 
attributes it to a Type~II QPT involving a jump between 
neutron number 58 and 60, from a weakly deformed 
A~configuration, to a strongly deformed B~configuration. 
Such a Type~II scenario is supported also by the trend in 
the magnetic moments ($\mu_J$) of the ground state, shown 
in~\cref{fig:order-params}(d), where both the data and the 
calculations show a constant value of $\mu_J$ for neutron 
numbers 52--58, and a drop to a lower value at 60, which 
persists for 60--64. This trend of approximately constant 
value for each range of neutron numbers, suggests a 
corresponding constant mixing in the ground state wave 
function, in line with the calculated weak mixing before 
and after the crossing, shown in~\cref{fig:order-params}(a).

To identify a \mbox{Type~I} QPT, involving shape changes  
within the intruder B~configuration, we examine the 
individual structure of Nb isotopes at the end points of 
the region considered. Figure~\ref{fig:93Nb-p} displays the 
experimental and calculated levels in $^{93}$Nb along with 
$E2$ and $M1$ transitions among them. The corresponding 
spectra of $^{92}$Zr, the even-even core, are also shown  
with an assignment of each level $L$ to the normal A or 
intruder B configurations, based on the analysis 
in Ref.~\cite{Gavrielov2022}, which also showed that the 
two configurations in $^{92}$Zr are spherical or 
weakly deformed. It has long been known~\cite{Heerden1973} 
that low-lying states of the A~configuration in $^{93}$Nb 
can be interpreted in a weak-coupling scheme, where the 
single-proton $\pi(1g_{9/2})$ state is coupled to 
spherical-vibrator states of the core. Specifically, for 
the $0^+_{1;\rm A}$ ground state of $^{92}$Zr, this 
coupling 
yields the ground state $9/2^+_1$ of $^{93}$Nb. For 
$2^+_{1;\rm A}$, it yields a quintuplet of states, 
$5/2^+_1, 7/2^+_1, 9/2^+_3, 11/2^+_1,13/2^+_1$, whose 
``center of  gravity'' (CoG)~\cite{Lawson1957}, is 
0.976~MeV, in agreement with the observed energy 0.935~MeV 
of $2^+_1$ in $\ce{^{92}Zr}$. The $E2$ transitions from the 
quintuplet states to the ground state are comparable in 
magnitude to the $2^+_{1;\rm A}\to0^+_{1;\rm A}$ transition 
in $\ce{^{92}Zr}$, except for $9/2^+_3$, whose decay is 
weaker. The corresponding $M1$ transitions are weak, while 
$M1$ transitions within states of the quintuplet are 
strong, as expected for weak-coupling to a spherical 
vibrator~\cite{IBFMBook}. An octet of states built on  
$4^+_{1;\rm A}$ can also be identified in the empirical 
spectrum of  $^{93}$Nb, with CoG of 1.591~MeV, close to 
1.495~MeV of $4^+_{1;\rm A}$.

The weak-coupling scenario is also valid for states of the 
intruder B configuration in $^{93}$Nb. As shown in 
Fig.~\ref{fig:93Nb-p}, the coupling of $\pi(1g_{9/2})$ to 
the $0^+_{2;\rm B}$ state in $^{92}$Zr, yields the excited 
$9/2^+_2$ state in $^{93}$Nb. For $2^+_{2;\rm B}$ it yields 
the quintuplet, $5/2^+_3, 7/2^+_3, 9/2^+_4, 11/2^+_3, 
13/2^+_2$, whose CoG is 1.705~MeV, a bit lower than 
1.847~MeV of $2^+_{2;\rm B}$. The observed $E2$ rates 
$1.03(9)$~W.u for \mbox{$9/2^+_2\to9/2^+_1$}, is close to 
the calculated value 0.85~W.u., but is smaller than the 
observed value $1.52(10)$~W.u for 
\mbox{$9/2^+_3\to9/2^+_1$}, suggesting that $9/2^+_2$ is 
associated with the B~configuration.

For $^{103}$Nb, the yrast states shown in \cref{fig:103Nb-p}
are arranged in a $K=5/2^+$ rotational band, with an 
established~\cite{Hotchkis1991} Nilsson model assignment 
$5/2^+[422]$. The band members can be interpreted in the 
strong-coupling scheme, where a particle is coupled to an 
axially deformed core. The indicated states are obtained by 
coupling the $\pi(1g_{9/2})$ state to the ground band 
($L=0^+_1,2^+_1,4^+_1,\ldots$) of $^{102}$Zr, which is  
associated with the intruder B~configuration. The 
calculations reproduce well the observed particle-rotor 
$J(J+1)$ splitting, as well as the $E2$ and $M1$ 
transitions within the band. 
Altogether, we see an evolution of structure from 
weak-coupling of a spherical shape in $^{93}$Nb, to 
strong-coupling of a deformed shape in $^{103}$Nb. Such 
shape-changes within the B~configuration (Type~I QPT), 
superimposed on abrupt configuration crossing (Type-II 
QPT), are the key defining feature of intertwined QPTs 
(IQPTs). Interestingly, the intricate IQPTs scenario, 
originally observed in the even-even Zr 
isotopes~\cite{Gavrielov2019,Gavrielov2022},
persists in the adjacent odd-even Nb isotopes.

In conclusion, we have presented a general framework 
(IBFM-CM), allowing a quantitative description of 
configuration mixing and related QPTs in odd-mass nuclei. 
An application to the positive-parity states in odd-even Nb 
isotopes disclosed a Type~II QPT (abrupt configuration 
crossing) accompanied by a Type~I QPT (gradual 
shape evolution and transition from weak to strong 
coupling within the intruder configuration), thus 
demonstrating, for the first time, IQPTs in odd-mass 
nuclei. The observed IQPTs in odd-$A$ Nb isotopes echo 
the multiple QPTs previously found in the adjacent  
even-even Zr isotopes \cite{Gavrielov2019, Gavrielov2022}. 
The results obtained motivate further experiments of 
non-yrast spectroscopy in such nuclei, as well as set the 
path for new investigations on multiple QPTs and 
coexistence in other Bose-Fermi systems.

\begin{acknowledgements}
This work is supported by the US-Israel Binational 
Science Foundation Grant No.~2016032. N.G. acknowledges 
support by the Israel Academy of Sciences of a Postdoctoral 
Fellowship Program in Nuclear Physics. We thank P.~Van 
Isacker for providing his IBFM code, which served as a 
basis for the IBFM-CM computer program.
\end{acknowledgements}


\begin{thebibliography}{99}
\bibitem{Gilmore1978a} 
R. Gilmore and D.~H. Feng, 
Phys. Lett. B \textbf{76}, 26 (1978).

\bibitem{Gilmore1979} 
R. Gilmore, 
J. Math. Phys. \textbf{20}, 891 (1979).

\bibitem{carr2010QPT} 
\textit{Understanding Quantum Phase Transitions}, 
edited by L. Carr (CRC, Boca Raton, FL, 2011).

\bibitem{Dieperink1980}
A.E.L. Dieperink, O. Scholten and F. Iachello,
Phys. Rev. Lett. \textbf{44}, 1747 (1980).

\bibitem{Cejnar2010} 
P. Cejnar, J. Jolie and R.~F. Casten, 
Rev. Mod. Phys. \textbf{82}, 2155 (2010).

\bibitem{Frank2006} 
A. Frank, P. Van Isacker and F. Iachello, 
Phys. Rev. C \textbf{73}, 061302(R) (2006).

\bibitem{Heyde11}
K.~Heyde and J.~L.~Wood,
Rev. Mod. Phys. \textbf{83}, 1467 (2011).

\bibitem{Gavrielov2019}
N. Gavrielov, A. Leviatan and F. Iachello,  
Phys. Rev. C \textbf{99}, 064324 (2019).

\bibitem{Gavrielov2022}
N. Gavrielov, A. Leviatan and F. Iachello,  
Phys. Rev. C \textbf{105}, 014305 (2022).

\bibitem{Casten2009}
R.F. Casten, 
Prog. Part. Nucl. Phys. \textbf{62}, 183 (2009).

\bibitem{Iachello2011}
F. Iachello,
Riv. Nuovo Cimento \textbf{34}, 617 (2011). 

\bibitem{Fortunato2021}
L. Fortunato,
Prog. Part. Nucl. Phys. \textbf{121}, 103891 (2021).

\bibitem{SM2005}
E. Caurier, G. Mart\'inez-Pinedo, F. Nowacki, A. Poves
and A. P. Zuker,
Rev. Mod. Phys. \textbf{77}, 427 (2005).
%The shell model as a unified view of nuclear structure

\bibitem{BMF2014}
B. Bally, B. Avez, M. Bender and P.-H. Heenen,
Phys. Rev. Lett. \textbf{113}, 162501 (2014).

\bibitem{ScholtenBlasi1982}
O. Scholten and N. Blasi,
Nucl. Phys. A \textbf{380}, 509 (1982).

\bibitem{IBFMBook}
F. Iachello and P. Van Isacker, 
\textit{The Interacting Boson-Fermion Model}  
(Cambridge University Press, Cambridge, U.K., 1991).

\bibitem{Petrellis2011a} 
D. Petrellis, A. Leviatan and F. Iachello,
Ann. Phys. \textbf{326}, 926 (2011).

\bibitem{Jolie2004}
  J. Jolie, S. Heinze, P. Van Isacker and R.F. Casten,
Phys. Rev. C \textbf{70}, 011305(R) (2004).

\bibitem{Alonso2005}
C.E. Alonso, J.M. Arias, L. Fortunato and A. Vitturi,
Phys. Rev. C \textbf{72}, 061302(R) (2005).

\bibitem{Alonso2007}
C.E. Alonso, J.M. Arias and A. Vitturi,
Phys. Rev. C \textbf{75}, 064316 (2007).

\bibitem{Alonso2009}
C.E. Alonso, J.M. Arias, L. Fortunato and A. Vitturi,
Phys. Rev. C \textbf{79}, 014306 (2009).

\bibitem{Boyukata2010}
M. B\"oy\"ukata, C.E. Alonso, J.M. Arias, L. Fortunato and
A. Vitturi,
Phys. Rev. C \textbf{82}, 014317 (2010).

\bibitem{Petrellis2011b} 
F. Iachello, A. Leviatan and D. Petrellis,
Phys. Lett. B \textbf{705}, 379 (2011).

\bibitem{Boyukata2021}
M. B\"oy\"ukata, C.E. Alonso, J.M. Arias, L. Fortunato and
A. Vitturi,
Symmetry \textbf{13}, 215 (2021).

\bibitem{Nomura2016a} 
K. Nomura, T. Nik\v si\'c and D. Vretenar,
Phys. Rev. C \textbf{93}, 054305 (2016).

\bibitem{Nomura2016b} 
K. Nomura, T. Nik\v si\'c and D. Vretenar,
Phys. Rev. C \textbf{94}, 064310 (2016).

\bibitem{Nomura2020} 
K. Nomura, T. Nik\v si\'c and D. Vretenar,
Phys. Rev. C \textbf{102}, 034315 (2020).

\bibitem{QuanMeng2018}
S. Quan, Z.~P. Li, D. Vretenar and J. Meng,
Phys. Rev. C \textbf{97}, 031301(R) (2018).

\bibitem{Spagnoletti2019}
P.~Spagnoletti, G.~Simpson, S.~Kisyov, D.~Bucurescu, 
J.-M.~R\'{e}gis, N.~Saed-Samii, A.~Blanc, M.~Jentschel, 
U.~K\"{o}ster, P.~Mutti,  T.~Soldner, G.~de~France,
C.~A.~Ur, W.~Urban, A.~M.~Bruce, C.~Bernards, F.~Drouet,
L.~M.~Fraile, L.~P.~Gaffney, D.~G.~Ghit\u{a},
S.~Ilieva, J.~Jolie, W.~Korten, T.~Kr\"{o}ll,
S.~Lalkovski, C.~Larijarni, R.~Lic\u{a}, H.~Mach, 
N.~M\u{a}rginean, V.~Paziy, Zs.~Podoly\'{a}k, P.~H.~Regan, 
M.~Scheck, J.~F.~Smith, G.~Thiamova, C.~Townsley, 
A.~Vancraeyenest, V.~Vedia, N.~Warr, V.~Werner and
M.~Zieli\'{n}ska, 
Phys. Rev. C \textbf{100}, 014311 (2019).

\bibitem{Boulay2020}
  F.~Boulay, G.~S.~Simpson, Y.~Ichikawa, S.~Kisyov,
  D.~Bucurescu, A.~Takamine, D.~S.~Ahn, K.~Asahi, H.~Baba,
  D.~L.~Balabanski, T.~Egami, T.~Fujita, N.~Fukuda,
  C.~Funayama, T. Furukawa, G.~Georgiev,A.~Gladkov,
  M.~Hass, K.~Imamura, N.~Inabe, Y.~Ishibashi,
  T.~Kawaguchi, T.~Kawamura, W.~Kim, Y.~Kobayashi,
  S.~Kojima, A.~Kusoglu, R.~Lozeva, S.~Momiyama, I.~Mukul, 
  M.~Niikura, H.~Nishibata, T.~Nishizaka, A.~Odahara,
  Y.~Ohtomo, D.~Ralet, T.~Sato, Y.~Shimizu, T.~Sumikama,
  H.~Suzuki, H.~Takeda, L.~C.~Tao, Y.~Togano, D.~Tominaga,
  H.~Ueno, H.~Yamazaki, X.~F.~Yang and J.~M.~Daugas, 
Phys. Rev. Lett. \textbf{124}, 112501 (2020).

\bibitem{Duval1981}
P.~D. Duval and B.~R. Barrett, 
Phys. Lett. B \textbf{100}, 223 (1981).

\bibitem{Duval1982}
P.~D. Duval and B.~R. Barrett, 
Nucl. Phys. A \textbf{376}, 213 (1982).

\bibitem{Sambataro1982} 
M. Sambataro and G. Moln\'ar, 
Nucl. Phys. A \textbf{376}, 201 (1982).

\bibitem{Ramos2014}
J.~E.~Garc\'\i a-Ramos and K.~Heyde,
Phys. Rev. C \textbf{89}, 014306 (2014).

\bibitem{Ramos2015}
J.~E.~Garc\'\i a-Ramos and K.~Heyde,
Phys. Rev. C \textbf{92}, 034309 (2015).

\bibitem{Nomura2016c}
K. Nomura, R. Rodr\'iguez-Guzm\'an and L. M. Robledo,
Phys. Rev. C \textbf{94}, 044314 (2016).

\bibitem{Lev2018}
  A. Leviatan, N. Gavrielov, J.~E.~Garc\'\i a-Ramos
  and P. Van Isacker,
Phys. Rev. C \textbf{98}, 031302(R) (2018).

\bibitem{Ramos2019}
J.~E.~Garc\'\i a-Ramos and K. Heyde,
Phys. Rev. C \textbf{100}, 044315 (2019).

\bibitem{Ramos2022}
E. Maya-Barbecho and J.~E.~Garc\'\i a-Ramos,
Phys. Rev. C \textbf{105}, 034341 (2022).

\bibitem{Barea2009}
J. Barea and F. Iachello, 
Phys. Rev. C \textbf{79}, 044301 (2009).

\bibitem{NDS.114.1293.2013}
C.M. Baglin,
Nucl. Data Sheets \textbf{114}, 1293 (2013).

\bibitem{NDS.112.1163.2011}
C.M. Baglin,
Nucl. Data Sheets \textbf{112}, 1163 (2011).

\bibitem{NDS.111.2555.2010}
S. Basu, G. Mukherjee and A Sonzogni,
Nucl. Data Sheets \textbf{111}, 2555 (2010).

\bibitem{NDS.111.525.2010}
N. Nica,
Nucl. Data Sheets \textbf{111}, 525 (2010).

\bibitem{NDS.145.25.2017}
E. Browne and J.Tuli,
Nucl. Data Sheets \textbf{145}, 25 (2017).


\bibitem{Guzman2011}
R. Rodr\'iguez-Guzm\'an, P. Sarriguren and L.~M. Robledo,
Phys. Rev. C \textbf{83}, 044307 (2011).

\bibitem{NDS.110.2081.2009}
D. De Frenne,
Nucl. Data Sheets \textbf{110}, 2081 (2009).



\bibitem{Cheal2009}
  B. Cheal, K. Baczynska, J. Billowes, P. Campbell,
  F. C. Charlwood, T. Eronen, D. H. Forest, A. Jokinen,
  T. Kessler, I. D. Moore, M. Reponen, S. Rothe,
  M. R\"uffer, A. Saastamoinen, G. Tungate, and
  J. \"Ayst\"o,
Phys. Rev. Lett. \textbf{102}, 222501 (2009).

\bibitem{Orce2010}
  J. N. Orce, J. D. Holt, A. Linnemann, C. J. McKay,
  C. Fransen, J. Jolie, T. T. S. Kuo, S. R. Lesher,
  M. T. McEllistrem, N. Pietralla, N. Warr, V. Werner,
  and S. W. Yates,
Phys. Rev. C \textbf{82}, 044317 (2010).

\bibitem{Heerden1973}
I.J. van Heerden, W.R. McMurray and R. Saayman,
Z. Physik \textbf{260}, 9 (1973).

\bibitem{Lawson1957}
R.D. Lawson and J.L. Uretsky,
Phys. Rev. \textbf{108}, 1300 (1957).

\bibitem{Hagen2017}
T. W. Hagen, A. G{\"{o}}rgen, W. Korten, 
L. Grente, M.-D. Salsac, F. Farget, 
I. Ragnarsson, T. Braunroth, B. Bruyneel, 
I. Celikovic, E. Cl{\'{e}}ment,G. de France, 
O. Delaune, A. Dewald, A. Dijon,
M. Hackstein, B. Jacquot, J. Litzinger,
J. Ljungvall, C. Louchart, C. Michelagnoli,
D. R. Napoli, F. Recchia, W. Rother,
E. Sahin, S. Siem, B. Sulignano,
Ch. Theisen and J.~J.~Valiente-Dobon,
Phys. Rev. C \textbf{95}, 034302 (2017).

\bibitem{Hotchkis1991}
M. A. C. Hotchkis, J. L. Durell, J. B. Fitzgerald,
A. S. Mowbray, W. R. Phillips, I. Ahmad, M. P. Carpenter,
R. V. F. Janssens, T. L. Khoo, E. F. Moore, L. R. Morss,
Ph. Benet and D.Ye,
Nucl. Phys. A \textbf{530}, 111 (1991).
\end{thebibliography}
\end{document}